\documentclass[dvips,rmp,twocolumn,groupedaddress,floatfix]{revtex4}
\usepackage{natbib}
\makeatletter
\renewcommand\@biblabel[1]{#1.}
\makeatother
\bibpunct{(}{)}{,}{n}{,}{,}
\usepackage[T1]{fontenc}
\usepackage{latexsym,dcolumn,graphicx,amsmath,amssymb,txfonts}
\usepackage{graphics}
\usepackage{graphicx}

\usepackage[dvips,dvipsnames]{color}
\definecolor{bgcolor}{rgb}{1,0.96,0.91}
\usepackage{url}
\definecolor{dark}{rgb}{0.6,0.0,0.0}
\usepackage[dvips,
breaklinks=true,
colorlinks=true,
citecolor=black,
pagecolor=dark,
linkcolor=dark,
menucolor=dark,%
urlcolor=dark,%
pdfborder={1 0 0}%
]{hyperref}%
\begin{document}

\title{Dynamics of Opinions and Social Structures}

\author{M. Rosvall}
\email{rosvall@u.washington.edu}
\affiliation{Department of Biology,
University of Washington,
Seattle, WA 98195-1800}
\author{K. Sneppen}
\email{sneppen@nbi.dk}
\affiliation{Niels Bohr Institute,
Blegdamsvej 17, Dk 2100, Copenhagen, Denmark}
\homepage{http://cmol.nbi.dk}
\date{\today}
\pacs{89.75.-k, 89.75.Fb, 89.70.+c}
\begin{abstract}
\textbf{Social groups with widely different music tastes, political
convictions, and religious beliefs emerge and disappear on scales
from extreme subcultures to mainstream mass-cultures.
Both the underlying social structure and the formation of opinions are dynamic and
changes in one affect the other.
Several positive feedback mechanisms have been
proposed to drive the diversity in social and economic systems
\citep{arthur,polya}, but little effort has been devoted to pinpoint
the interplay between a dynamically changing social network and the
spread and gathering of information on the network.
Here we analyze this phenomenon in terms of a social network-model that explicitly
simulates the feedback between information assembly and emergence of
social structures: changing beliefs are coupled to changing relationships
because agents self-organize a dynamic network
to facilitate their hunter-gatherer behavior in
information space. Our analysis demonstrates that tribal organizations and
modular social networks can emerge as a result of contact-seeking agents
that reinforce their beliefs among like-minded. We also find that
prestigious persons can streamline the social network into hierarchical
structures around themselves.}
\end{abstract}
\maketitle

The competition between segregation and coherence in social systems has
long been a subject of both practical and theoretical interest.
T.\ Schelling proposed simple models to understand
how segregation emerge in urban areas \citep{schelling} and B.\ Arthur
suggested that the emergence of industrial centers
is a result of positive feedback between agencies that prefer to
be close to similar agencies \citep{arthur}.
In general, the goal of individuals to understand and agree
with their closest associates \citep{lazarsfeld,mcpherson,zeggelink}
can be obtained by either merging opinions with friends
or by creating new contacts. In this spirit, we will discuss a
social network model where individuals can use their connections to
communicate and update their views with friends
or establish new contacts via friends to shortcut communication pathways.
By modeling communication on the network and adaptive rewiring of the network
together, we reveal strong reinforcement of opinions in emerging group structures.

\begin{figure}[t!]
\centering
\includegraphics[width=1.0\columnwidth]{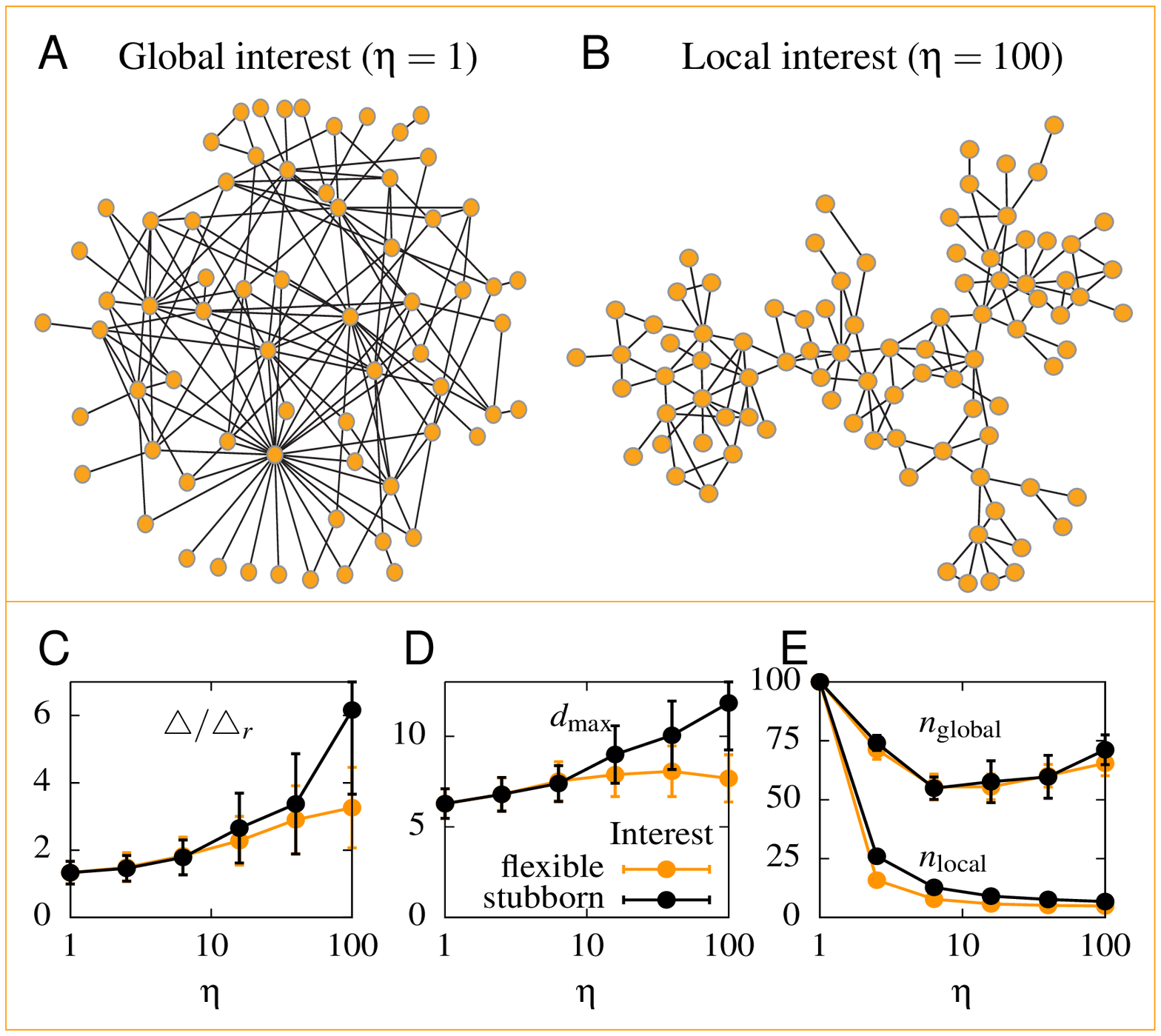}
\caption{\small Interplay between network structure and
the extent $\eta$ to which recent gossiping influences the priority in social climbing.
Varying $\eta$, the
bottom pane illustrates cliquishness (C), diameter (D), and social
horizon (E).
Simulations based on $C=10$ communication events per link for each rewiring event in the system with system size
fixed to $N=100$ and $L=150$ links.
The results are robust to a hundredfold drop in the communication to rewiring ratio,
but break down at an even lower communication rate when only small groups can be maintained by the communication.
Panels C-E also illustrate the dependence of the rate of opinion adaptation, flexibility,
with stubborn adaptation referring to a $\mu=0.01\%$ change of the interest elements per communication event,
and flexible referring to $\mu=1\%$ change.
We define the flexibility as $\mu C L$.
Stubborn adaptation  corresponds to a flexibility of 15\% changes in all interest memory when all links are changed once
whereas flexible adaptation corresponds to complete reallocation.
\label{fig1}}
\end{figure}

To comprehend the dynamics of opinions in a social system,
it is important to understand the coupling to the social structure the dynamics is embedded in \cite{zeggelink,carpenter,heclo,bala}.
For example, if anyone shares information with anyone else without any preference, the society can be described by a simple mean-field model
where everyone has access to all information.
On the other hand, if people only interact
with a small number of friends, the description of the system
must reflect this communication constraint.
A network representation of the interactions in the system
is a powerful way to depict the limited access to information;
a static network with exchange of information between and about nearest neighbors alone
represents a scenario with the most limited access to global information. Which world do we live in?

Taking the latter perspective, an immense literature on 
opinion formation on static networks,
most straightforward formulated in the voter 
model \cite{liggett,redner}, has helped us understand how the structural constraints affect the dynamics.
In this letter, we take a different approach and instead ask
where the preference for specific information comes from
and how the network constraints can be overcome to make this information accessible.

To open up the world beyond nearest neighbors, we 
present a game in which we 
let agents improve their positions in a dynamic network
based on their limited local perception of the system.
To maintain this perception, they
continuously communicate to obtain (1) a picture of where other agents are in the network and (2) an opinion about the importance of other
agents. We incorporate information flows from distant parts
of the social network into the model by letting two connected agents communicate and share information about any third agent in the network.
This indirect information-gathering
resembles the frequent gossiping in daily life
and its consequences for non-local information assembly has been
investigated in static social networks \citep{friedkin-infoflow}.

To incorporate the above elements in a simple model,
we give each agent an individual memory. The memory includes three vectors with
(1a) $N$ pointers associated to the $N$ agents in the system that show which friend provided
information about the agent, (1b) the age of each of the $N$ pieces of information to compare the quality of the information with friends,
and (2) the names of other agents filling the memory to
an extent that reflects the interest in these agents.
The memory in (1a) and (1b) is a map and (2) is the priority.

We emphasize that all agents have a pointer, an age of this pointer
and a priority for each other agent in the system.
The pointer and its age in itself is enough to
allow communicating agents to build an operative map \cite{rosvall1} 
of their own position in the network as described below.
With the priority memory, we can incorporate
simplified human behavior by introducing rules that mimic
individuals' choices when selecting who they want
to connect to or talk about. This makes it possible to
explore the mechanisms behind the emergence 
opinions among individuals.

The memory is updated through
communication over links in the network, which in turn are rewired
based on the locally obtained knowledge.
Because of the ongoing changes in the social network,
the obtained knowledge becomes outdated with time. 
Accordingly, when two agents communicate about a third agent,
they first decide which of them that has the newest information \citep{rosvall1}.
This information is considered the most reliable,
and the agent with the older information updates 
its information using
the agent with the newer information as a reliable source of 
information about the third agent. 
We increment the age in (1b) for all agents when every 
link has participated in on average one communication event in the system.
Because every agent always has information with age 0 about itself,
the age of the information about an agent will tend to get older
the further away it is from the agent in the network.
The age of the information is therefore a good proxy for its quality,
assuming that agents are not lying.

The basic model, accessible as a Java
applet with a detailed description \citep{java}, is defined in terms of $N$ agents with a fixed
number $L$ of links. The network model is executed in time steps, each
consisting of one of the two events:
\begin{itemize}
\item
\emph{Communication $C$:} Chose a random link and let the two
agents connected by the link communicate about a third agent
\emph{selected} by one of them. The two agents also update their
information about each other.
\item
\emph{Social climbing $R$:} Let a random agent use the local information to
form a link to a friend's friend to shorten its distance to a \emph{selected} third
agent. Subsequently a random agent loses one of its links.
\end{itemize}

Communication or gossiping 
involves evaluation of information
quality based on its age as described above.
The social climbing, which corresponds to rewiring of the network,
is a slow process compared to gossiping.
If this was not the case, random people would share reliable information
with anybody and the interactions could be simpler described by a mean-field
model. We therefore simulated the model with 10 communications per link for each
rewiring event in the system. Links are formed to friends of friends
similar to triadic closure \citep{rapoport}, reflecting a
gradual social climbing. Here friends refer to the particular
agents that have provided the most recent information about the selected agent.
The new links are therefore formed on basis of the memory
rather than on basis of the present network
\citep{fragment}.

\begin{figure}
\centering
\includegraphics[width=1.0\columnwidth]{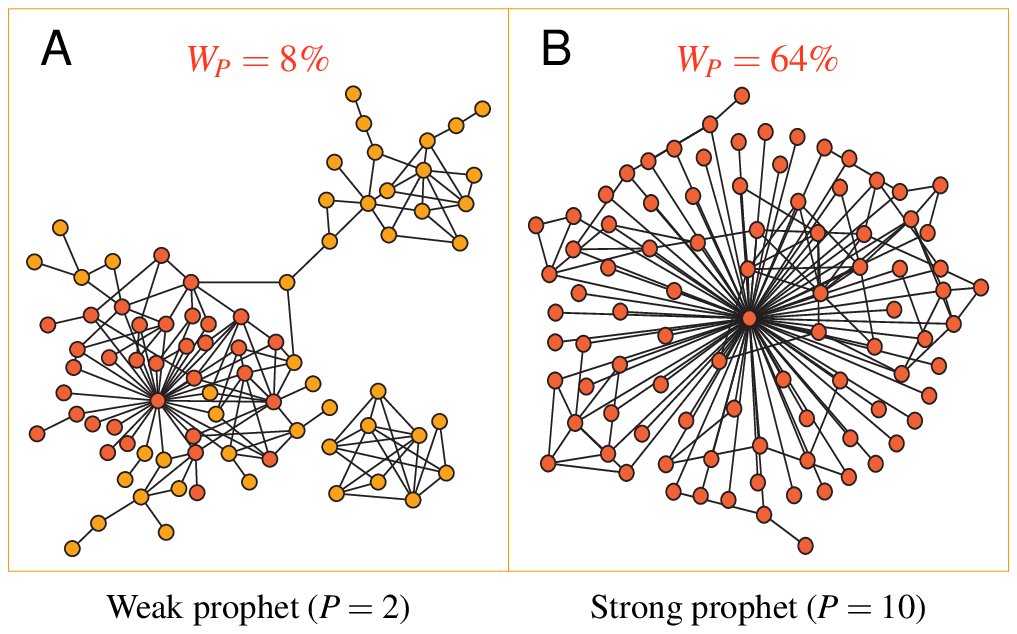}
\caption{\small Network structures where one agent, the prophet
P, has increased prestigious power to capture an unusually large
fraction of the memory of whoever she talks to. The results are
based on a prophet with (A) weak prestige $P=2$, and (B) high prestige
$P=10$. Note that when sub-networks
separate entirely, as in panel A, the agents still maintain information about
who they should contact in the social climbing step and
separated clusters can reconnect. The $W_P$ values
correspond to the fraction of global interest devoted to the prophet. Other
parameters as in Fig.~\ref{fig1}B. \label{fig2}}
\end{figure}

To be able to investigate the reinforcement of opinions,
we vary the way we specify the selection criterion for the third person, the agent of interest,
in the communication and the social climbing events.
Figure \ref{fig1}A and B illustrate that two different selection mechanisms
can give rise to widely different social networks.
To generate the network in Fig.~\ref{fig1}A, we let the agents select
the agent of interest randomly with equal chance for any agent.
In contrast, to generate the modular network in Fig.~\ref{fig1}B, we let
one of the communicating agents use her prioritized memory to
select the agent of interest (in a communication event we always let the other
agent passively accepts the choice).
Following Spencer's \citep{spencer} observation of
proportionality between interest and previous experience,
an agent's priority memory is filled with other agents' names
proportional to their occurrence in the agent's resent gossiping.
Related to this use of proportionate selection is the work by H.~Simon \citep{simon} to
explain Zipf's law and the work presented in refs.~\citep{yasutomi,donangelo} to model
emergence of fashions.

The priority memory is essential for modular structures to emerge.
Let us therefore describe in more detail how we implement
the proportionate selection when building the priority memory
for each agent. We regulate the extent to which agents select the topic of conversation
according to recent gossiping by the external parameter $\eta$, which may also be understood as the ratio of local to global interest.
For each agent $j$, we model the
priority memory in (2) above
by $N\eta$ elements. The first $N$ elements are static and fixed to each of the $N$ agents' names corresponding to the global interest.
The remaining memory can be changed by communication.
When agent $j$ talks to, or hears about,
another agent $i$, the name of $i$ will randomly replace a fraction $\mu$ of this memory. 
Old priorities will thereby slowly fade as they are replaced
by new topics of interest.

In selecting communication topic
or aim of social climbing, agent $i$ is chosen with probability $w_j(i\,)=n_j(i\,)/(N\eta )$
proportional to how many times the name of $i$ occurs in agent $j$'s
priority memory, $n_j(i\,)$.
The size of $\eta$
determines the degree of preferential allocation of memory.
For $\eta=1$ any topic is selected with equal chance (global interest, see Fig.~\ref{fig1}A),
whereas larger $\eta$ increases the chance that the agents
choose a topic proportionate to personal experience \cite{spencer} (local interest, see Fig.~\ref{fig1}B).

A dominant local interest with large $\eta$ predicts an evolving network with pronounced modular structure as in Fig.~\ref{fig1}B. The emerging modular network does no longer develop large hubs, but instead shows large cliquishness
quantified by the number of triangle motifs in units of the random expectation \citep{foot2},
$\bigtriangleup /\bigtriangleup_{r}$ (see
Fig.~\ref{fig1}C). Figure \ref{fig1}D shows that the diameter
$d_{\mathrm{max}}$ of the network easily doubles as $\eta$ increases,
weakening the small-world effect \citep{watts}
and the global navigability of the network \citep{kleinberg} (to illustrate the robustness of the model,
we in Fig.~\ref{fig1}C-E show the results for two different opinion adaptation rates $\mu$).

To quantify the locally acquired interest we counted the effective
number of persons, $n_{\mathrm{local}}$, a typical agent has in
the interest memory
\begin{equation}
n_{\mathrm{local}} \; = \; \langle \frac{1}{N\langle w_{j}^{2}(i\,) \rangle}_i
\rangle_j.
\end{equation}
Here $\langle w_{j}^{2}(i\,) \rangle_i$ averages over the
weights allocated to different interests $i$ of agent $j$.
Figure \ref{fig1}E shows this social horizon of the individual
agent, $n_{\mathrm{local}}$, as well as the global horizon where all
memories are pooled together in $n_{\mathrm{global}}$ \citep{foot3}.
As $\eta$ increases, $n_{\mathrm{local}}$ collapses while
$n_{\mathrm{global}}$ remains on the scale of $N$ --- the development
toward social cliques is democratic, with anyone getting a fair
share of attention while still allowing people to focus locally on
members of their particular ``club''.
The extent to which this club is maintained and closed for
migration is controlled by the flexibility defined in the caption of Fig.~\ref{fig1}; the modular structure
breaks down when one changes opinion faster than one changes friends.
\bigskip

We would like to emphasize that our model presents
a \emph{spherical cow} of human
memory and response quantified through 3 parameters:
the communication level $C$, the local to global interest $\eta$, and the flexibility $\mu$. For $C$ sufficiently high, our
results are independent of further increase in
communication. At any reasonably low level of flexibility
$\mu$, the evolving system develops a modular structure
and a diminished information horizon with increased
$\eta$. Thus the model emphasizes the proportionate
interest allocation \cite{spencer,simon}
as the key parameter for development of segregation.

Our idealized model-world consists of agents with equal properties.
In spite of this equality, the model predicts
segregation in the form of a social network with modular
structure with widely different priorities and opinions.
The local agreement and global divergence
self-organize as a consequence of repeated recent communication and
reinforced contacts to people one gossips about.

Central to the model is to build and use the priority
memory. Here we have explored a particularly simple linear
model for both the construction and the use of priorities.
But we stress that the model framework can be extended to more detailed networking games, 
including for example trust \cite{skyrms}, cheating agents \cite{rosvall}, or
update of priorities based on experiences of the reliability of the
obtained information. Undoubtedly, real humans will have different intrinsic properties.
As the simplest example of how heterogeneous agents may influence the
system, we will study the case where one agent has one special property.

We consider a simple directed strategy to
influence public opinion aimed to increase the status of a
particular agent.
\begin{itemize}
\item
\emph{Prophet} P: each time agent P communicates, she imposes
her personality on the agent $j$ she talks to by converting a fraction $P\mu$ of this agent's interest memory to P.
\end{itemize}
The ``prophet'' uses the local network structure in analogy to the
way the so called ``heroes'' or social leaders have influenced
the society with their acquired prestige \citep{carlyle,goode,henrich}.

The spreading of information across the system is analogous to
viral marketing \citep{brown}, which makes it possible for the prophet to reinforce her
position by gaining visibility and subsequently links from larger
parts of the system. Moreover, higher connectivity makes the prophet
preferentially more accessible to people and the prestige-biased
transmission of ideas \cite{henrich} modeled by the priority memory connects the positive feedback.
When $P=1$ the prophet P
becomes a normal agent, and the overall network develops to resemble
the one in Fig.~\ref{fig1}B. Figure \ref{fig2} shows that an
agent's increased ability to positively bias other persons'
interest memories drives the network topology toward a centralized
structure. For example, a single prophet with strength $P=2$ can generate a substantial group of followers (Fig.~\ref{fig2}A). The figure also illustrates that
separated social groups can emerge and collapse by temporarily weaker
interest in other agents. Figure \ref{fig2}B shows that a strong
prophet drives the full system to a single hierarchical
structure. In fact, by measuring the social horizons we find that a single P
with strength $P=10$ drives both $n_{\mathrm{local}}$ and
$n_{\mathrm{global}}$ to about $2$ --- a dominating prophet can
efficiently initiate a totalitarian system to an extent that people
barely think about anybody except the social leader.
This facilitates social coherence and
order, but obviously not the formation of diverse cultures or
coexistence of small communities.
\\

In general, the emerging structures are robust consequences of an interplay between
the following positive feedback mechanisms:

\noindent
\begin{center}
\includegraphics[width=1.0\columnwidth]{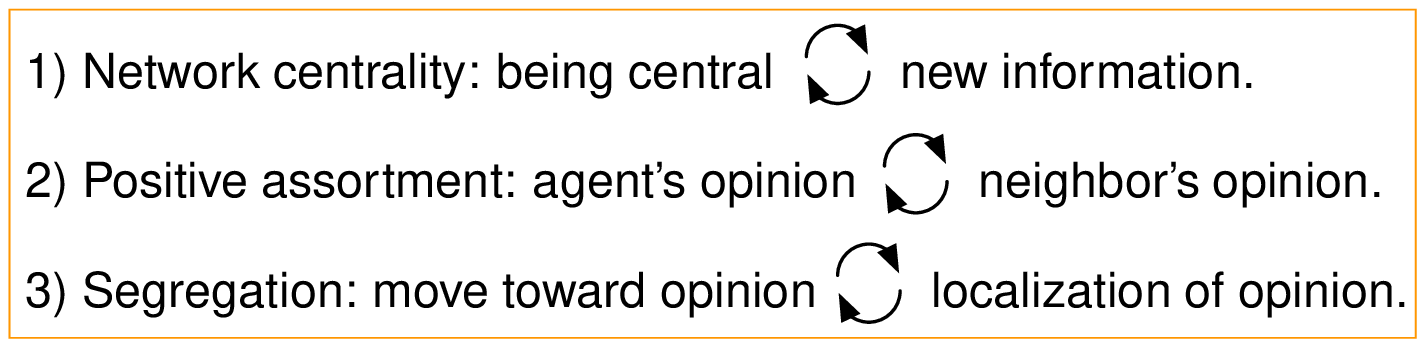}\\
\end{center}
Without individuals with personal interests only the
first feedback is active, but is in itself enough to give the
network a broad degree-distribution \citep{rosvall}. The two
subsequent reinforcements generate interest groups and segregation
manifested in a modular network.
Together these feedbacks make it very favorable
to manipulate opinion spreading. Here we have presented
a framework which can give insight into the mechanisms
of communication strategies in historical as well as modern propaganda.
\bigskip

\begin{acknowledgments}
We would like to acknowledge Anders Lisdorf for constructive comments on the manuscript.
This work was supported by the Danish National Research Foundation
for support through the center for Models of Life.
MR was also supported by National Institute of General Medical Sciences Models
of Infectious Disease Agent Study Program Cooperative Agreement
5U01GM07649.
\end{acknowledgments}


\begin{thebibliography}{10}
\bibitem{fragment}
Technically the communication events are limited to current network, whereas
  formation of new links rely on memory of links that may not exist any longer.
  Thereby even disconnected nodes will eventually reconnect.
\bibitem{foot2}
$\bigtriangleup$ is the number of triangles in the model network, whereas
  $\bigtriangleup_{\mathrm{random}}$ is average number of triangles in a
  randomized network with exactly the same degree of each node as the model
  network.
\bibitem{foot3}
Vieving memories of all agent as one common pool, we count the total weights of
  all memory slots allocated to agent $i$ as $W(i\,)=\sum_j w_j(i\,) \in [0;N]$
  and define the global information measure for the effective agent diversity
  as $n_{global}= N/\langle W^2(i\,) \rangle_i$.
\end{thebibliography}
\end{document}